\documentstyle{article}

\author{L. S. F. Olavo\\
Departamento de Fisica - Universidade de Brasilia - UnB\\
70910-900 - Brasilia - D.F. - Brazil}
\title{Quantum Mechanics as a Classical Theory XI: \\
Thermodynamics and Equilibrium
}

\begin{document}

\maketitle
\begin{abstract}
In this continuation paper the theory is further extended to reveal the
connection between its formal aparatus, dealing with microscopic quantities,
and the formal aparatus of thermodynamics, related to macroscopic properties
of large systems. We will also derive the Born-Sommerfeld quantization rules
from the formalism of the infinitesimal Wigner-Moyal transformations and, as
a consequence of this result, we will also make a connection between the
later and the path integral approach of Feynman. Some insights of the
relation between quantum mechanics and equilibrium states will be given as a
natural development of the interpretation of the above results.
\end{abstract}

\section{Introduction}

We have already deserved much of our attention to the microscopic behavior
of systems described by the Schr\"odinger equation, which we derived from
Liouville's equation together with the classical dynamical equations\cite
{eu1} -\cite{eu10}.

It remains now to try to establish the connection between this microscopic
behavior and those macroscopic, which manifest themselves when the systems
comprising the {\it ensemble} have a sufficiently large number of particles.

In the second and third sections of this paper we will make the connection
between the microscopic quantities, derived from the quantum formalism, and
the usual thermodynamics quantities (entropy, free energy, etc.) using the
canonical {\it ensemble} model.

As everybody knows, in this approach we allow the systems ($S$) composing
the {\it ensemble} to interact with a neighborhood ($O$), usually called the
heat bath. In such cases, the interaction is still considered sufficiently
feeble as to allow one to write a hamiltonian function $H(q,p)$ for $S$ not
depending on the degrees of freedom of $O$. The system $O$ is necessary only
as a means of imposing its temperature $T$ upon system $S$.

\section{Equilibrium and Quantum Formalism}

According to Gibbs, in a state of equilibrium we shall have a canonical
probability distribution defined as 
\begin{equation}
\label{F1} F(q,p)=Ce^{-2\beta H(q,p)}, 
\end{equation}
where 
\begin{equation}
\label{F2} 2\beta=\frac{1}{K_B T}, 
\end{equation}
with $K_B$ being the Boltzmann constant, $T$ the absolute temperature and $C$
some normalization constant.

The hamiltonian function may be written as 
\begin{equation}
\label{F3} H(q,p)=\sum_{n=1}^{N}{\frac{p_n^2}{2m_n}}+V(q_1,..,q_N), 
\end{equation}
where we are supposing systems $S$ with $N$ degrees of freedom with
potential function not depending on velocities nor time.

Using the infinitesimal Wigner-Moyal transformation 
\begin{equation}
\label{F3a} \rho(q-\delta q/2,q+\delta q/2)=C\int{\ F(q,p)e^{i\sum{p_n\delta
q_n/\hbar} }dp }, 
\end{equation}
the characteristic function becomes 
\begin{equation}
\label{F4} \rho(q-\delta q/2,q+\delta q/2)=C\int{\ e^{-2\beta\left( \sum{%
p_n^2/2m_n}+V(q_1,..,q_N)\right)} \cdot e^{i\sum{p_n \delta q_n}/\hbar} }%
dp_1..dp_n 
\end{equation}
which gives, performing the integral, 
\begin{equation}
\label{F5} \rho(q-\delta q/2,q+\delta q/2)=C_1 e^{-2\beta V(q_1,..,q_N)}
e^{-\sum{\frac{m_n}{4 \beta \hbar^2}(\delta q_n)^2} }. 
\end{equation}

Clearly, this characteristic function is a solution to the equation 
\begin{equation}
\label{F6} -\sum_{n=1}^N \frac{\hbar^2}{m_n}\frac{\partial^2 \rho} {\partial
q_n \partial (\delta q_n)} +\sum_{n=1}^N {\frac{\partial \rho}{\partial q_n}%
(\delta q_n) \rho}=0, 
\end{equation}
obtained from the Liouville equation performing the infinitesimal
transformation (\ref{F3a}) upon it.

As was already discussed \cite{eu1}, it may be possible to write the
characteristic function in the format 
\begin{equation}
\label{F9} \rho(q-\delta q/2,q+\delta q/2)=\psi^{\dag}(q-\delta
q/2;t)\psi(q+\delta 2/2;t) 
\end{equation}
to be able to derive the Schr\"odinger equation related with the problem for
the amplitudes. In this case, it is easy to see that the amplitudes $\psi$
may be written as 
\begin{equation}
\label{F10} \psi(q;t)=\sqrt{C_3}e^{-\beta V(q_1,..,q_N)}e^{-iEt}, 
\end{equation}
giving, for the characteristic function%
\footnote{As is expected, the
terms having order greater than or equal to 3 in $\delta q$ shall not
be considered since this variable is considered infinitesimal. Its
second power shall be maintained because of the derivative
$\partial/\partial (\delta q)$ we have in equation (\ref{F6}) above
for the characteristic function.} 
\begin{equation}
\label{F11} \rho(q-\delta q/2,q+\delta q/2)=C_3
e^{-2\beta\left[V(q_1,..,q_N)+ \frac{1}{8}\sum{(\delta q_n)^2\frac{%
\partial^2 V}{\partial q_n^2} }\right]} . 
\end{equation}

Comparing the expression (\ref{F11}) with (\ref{F5}) we observe that it is
necessary to have, around the point $q=(q_1,..,q_n)$ where the function is
being evaluated 
\begin{equation}
\label{F14}\left. \frac{\partial ^2V}{\partial q_n^2}\right| _{\delta q_n=0}=
\frac{m_n}{\beta ^2\hbar ^2},
\end{equation}
if we want to write the characteristic function as in (8) with the
amplitudes given in (9).

Expression (\ref{F11}) is equivalent to take 
\begin{equation}
\label{F15} \rho_{eq}(q)=e^{-2\beta V(q_1,..,q_N)} 
\end{equation}
as the probability density function for the thermodynamic equilibrium,
defined upon configuration space, and express the characteristic function as 
\begin{equation}
\label{F16} \rho(q-\delta q/2,q+\delta q/2)=\rho_{eq}(q+\delta
q/2)=\rho_{eq}(q-\delta q/2) 
\end{equation}
{\it if we have} 
\begin{equation}
\label{F17} \left. \frac{\partial V}{\partial q_n} \right|_{\delta q_n=0}=0 
\end{equation}
at the considered point. The point to which we have both (11) and (14)
satisfied defines, as known, a mechanical equilibrium point for the considered
physical system.

This means that the characteristic function, for this specific problem where
we consider an {\it ensemble} of systems ($S$) in thermal equilibrium with a
reservoir $O$, can be considered as the probability density function when
evaluated at points infinitesimally distant from the systems' mechanical
equilibrium situations.

Obviously, the density $\rho_{eq}(q)$ is the function obtained from the
characteristic function (\ref{F5}) taking the limit $\delta q_n,n=1,.., N$%
\footnote{Indeed, it will be also correct to use this property to get
the functional description of the probability amplitudes, as given by
equation (\ref{F10}); this, of course, neglecting the phase factor that
remains ambiguous.} or, what is equivalent, performing the integration of
the probability density function defined on phase-space (\ref{F1}) with
respect to the variables $p_n,n=1,..,N$. This is also consistent with the
expression\footnote{See
last note.} 
\begin{equation}
\label{F18} \rho_{eq}(q)=\psi^{\dag}(q;t)\psi(q;t), 
\end{equation}
as expected.

The development above furnishes a very well-defined physical interpretation
for the characteristic function:

{\it The characteristic function is obtained from the probability density,
the later being taken at infinitesimally distant points from the mechanical
equilibrium situation of the system.}

This connection between the separation (\ref{F9}) of the characteristic
function---that allows the very derivation of the Schr\"odinger
equation---and the fact that we are dealing with systems infinitesimally
near the mechanical equilibrium points, make one recall the physicists first
intuitions at the beginning of this century. The above conclusions, although
derived for a particular problem, gives us a first insight on the validity
of the Bohr
postulates as they were first formulated in the early days of quantum
formalism\footnote{In the next section this connection will be made clear.}.

We may go further and ask which Schr\"odinger equation is related with the
amplitude (\ref{F10}). In this case, substituting this amplitude into the
Schr\"odinger equation 
\begin{equation}
\label{F19}-\sum_{n=1}^N{\frac{\hbar ^2}{2m_n}\frac{\partial ^2\psi }{%
\partial q_n^2}}+V(q_1,..,q_N)\psi =E\psi ,
\end{equation}
we get 
\begin{equation}
\label{F20}\sum_{n=1}^N{\frac{\beta \hbar ^2}{2m_n}\frac{\partial ^2V}{%
\partial q_n^2}}+V(q_1,..,q_N)-\sum_{n=1}^N{\frac{\beta ^2\hbar ^2}{2m_n}%
\left[ \frac{\partial V}{\partial q_n}\right] ^2}=E
\end{equation}
from which we may write, using equations (\ref{F14}) and (\ref{F17}) 
\begin{equation}
\label{F21}E=V(q_1^0,..,q_N^0)+NK_BT,
\end{equation}
where $q_n^0$ represents the mechanical equilibrium point related with this
degree of freedom. The second term on the right-hand side of equation (\ref
{F21}) represents the energy related with the reservoir $O$; in such case
this reservoir may be interpreted as consisting of $N$ independent harmonic
oscillators, each one contributing, as prescribed by the equipartion theorem
with the energy $K_BT$%
\footnote{Note that we are using the Boltzmann (\ref{F1})
`classical' distribution function, for which this result aplies.}.

It is interesting to note, from expression (\ref{F5}) for the characteristic
function, seen as obtained from the density $\rho _{eq}(q)$ that, for the
cases where the absolute temperature tends to zero, any departure $\delta q$
from zero, whatever small, gives origin to an extremely small density of
states, or, as implied, it is extremely unprobable to find the systems out
of this mechanical equilibrium point, for this temperature. Because of
equation (\ref{F21}) we have 
\begin{equation}
\label{F21a}E=V(q_1^0,..,q_N^0),
\end{equation}
and it is as if the reservoir does not exist.

\section{Connection with Thermodynamics}

As a consequence of the approach of the last section, it is possible to
establish a connection between the microscopic entities of the quantum
formalism and the macroscopic description given by thermodynamics.

To reach this goal we may define the free energy 
\begin{equation}
\label{F22} F_G(q_1,..,q_N)=V(q_1,..,q_N), 
\end{equation}
such that 
\begin{equation}
\label{F23} F_G=-K_B T ln(\psi^{\dag}(q)\psi(q)). 
\end{equation}

Expressing the entropy function as 
\begin{equation}
\label{F24} S=K_B ln(\psi^{\dag}(q)\psi(q)), 
\end{equation}
we find 
\begin{equation}
\label{F25} F_G=-TS. 
\end{equation}

Equation (\ref{F24}) represents the desired connection between microscopic
properties, described by the probability amplitudes satisfying a
Schr\"odinger equation, and the macroscopic behavior of the system,
described by the entropy and its related functions.

\section{The Bohr-Sommerfeld Rules}

It seems curious, at first sight, that in equation (\ref{F16}) the
characteristic function may be interpreted, at least in this particular
case, as the probability density in configuration space taken at a point $%
\delta q/2$ apart from the equilibrium situation---and not $\delta q$ as one
might think.

Indeed, if we write 
\begin{equation}
\label{F21.1} \rho(q-\delta q/2,q+\delta q/2;t)=\int{e^{\frac{i}{\hbar}%
p\delta q}F(q,p;t)dp} 
\end{equation}
and interpret $p$ as $p=-i\hbar\partial/\partial q$, then we end with the
formal identification 
\begin{equation}
\label{F21.2} \rho(q+\delta q;t)=\int{e^{\delta q\frac{\partial}{\partial q}%
}F(q,p;t)dp}. 
\end{equation}

Such an approach would be, however, misleading. As was already shown \cite
{eu1}-\cite{eu10}, the function $p$ is taken into the above cited operator
only when it is acting upon the probability amplitudes and not when it is
acting upon the density function, as in equation (\ref{F21.1}).

We may go to a representation of $F(q,p;t)$, based on phase space
probability amplitudes, imposing that we may write it from a function 
\begin{equation}
\label{F21.3} f(q,p;p^{\prime};t)=\phi^{\dag}(q,2p-p^{\prime};t)\phi(q,p^{%
\prime};t), 
\end{equation}
such that 
\begin{equation}
\label{F21.4} F(q,p;t)=\int_{-\infty}^{+\infty}{f(q,p;p^{\prime};t)dp^{%
\prime}}= \int_{-\infty}^{+\infty}{\phi^{\dag}(q,2p-p^{\prime};t)\phi(q,p^{%
\prime};t)dp^{\prime}}. 
\end{equation}
In this case, the integration in (\ref{F21.1}) becomes, using the
convolution theorem, given by 
\begin{equation}
\label{F21.5} \rho(q-\delta q/2,q+\delta q/2;t)=T_{F}\{ \phi^{\dag}(q,p;t)
\} T_{F}\{ \phi(q,p^{\prime};t) \}, 
\end{equation}
where $T_{F}\{ \phi \}$ represents the Fourier transform of the function $%
\phi$.

Now, writing 
\begin{equation}
\label{F21.6} \psi(q+\delta q/2;t)=T_{F}\{ \phi(q,p;t) \}=\int{e^{\frac{i}{%
2\hbar}p\delta q} \phi(q,p;t)dp}, 
\end{equation}
such that 
\begin{equation}
\label{F21.7} \psi^{\dag}(q-\delta q/2;t)=T_{F}\{ \phi(q,p;t) \}=\int{e^{
\frac{i}{2\hbar} p\delta q} \phi^{\dag}(q,p;t)dp}, 
\end{equation}
we reach%
\footnote{Note that the equations (\ref{F21.6}) and (\ref{F21.7}) are
compatible with the formal identification $p=-i\hbar\partial/\partial q$.,
since
$$
\psi(q+\delta q/2;t)=\int{e^{\frac{\delta q}{2}\frac{\partial}{\partial q}}
\phi(q,p;t)dp}=\int{\phi(q+\delta q/2,p;t)dp},
$$
giving the correct displacement we found for the density function.}, in
expression (\ref{F21.5}), 
\begin{equation}
\label{F21.8} \rho(q-\delta q/2,q+\delta q/2;t)=\psi^{\dag}(q-\delta q/2;t)
\psi(q+\delta q/2;t), 
\end{equation}
as was previously \cite{eu1} imposed%
\footnote{ 
The constraint given by equations (\ref{F21.3}) and (\ref{F21.4}) is
similar, therefore, to the imposition that we have done directly upon
the characteristic function (\ref{F21.8}) in other papers.}.

The expressions (\ref{F21.6}) and (\ref{F21.7}) may be used to make the
bridge between this formalism and the one proposed by Bohr and latter
deepened by Sommerfeld, known as the `old quantum theory'.

In this case, considering expression (\ref{F21.6}), for example, we may note
that it consists in an integral transformation with a nucleus given by%
\footnote{We wrote $p(q)$ to make it clear that, in the phase space, the 
system follows a trajectory such that this functional identification is
allways possible. Indeed, considering a conservative system, we may
write the hamiltonian as $H(q,p)=\alpha$, where $\alpha$ is a constant,
and obtain $p$ as a function of $q$\cite{Goldstein1}.} 
\begin{equation}
\label{F21.9} K_{p(q)}(q+\delta q,q)=e^{\frac{i}{\hbar}p(q)\delta q}. 
\end{equation}

If the system is {\it periodic}---independently of being a libration or a
rotation \cite{Goldstein1}---then we may perform a succession of
transformations taking 
\begin{equation}
\label{F21.10} \psi(q+Q;t)=\pm \psi(q;t), 
\end{equation}
where $Q$ is the period (in q-space) of the movement and the signal $\pm$
takes into account the fact that we are dealing with amplitudes and not with
densities, leaving the choice of the signs undefined.

The nucleus of this transformation is given by 
\begin{equation}
\label{F21.11} \psi(q+Q;t)=\int{\ K_{p(q)}(q+Q,q)\phi(q,p;t)dp}, 
\end{equation}
such that 
\begin{equation}
\label{F21.12} K_{p(q)}(q+Q,q)=\lim_{N\rightarrow \infty}\prod_{n=1}^{N}
K_{p(q+(n-1)\delta q)}(q+n\delta q,q+(n-1)\delta q), 
\end{equation}
where we made 
\begin{equation}
\label{F21.13} N\delta q=Q, 
\end{equation}
being necessary to take the limit $N\rightarrow \infty$ since $\delta q$ is
infinitesimal.

Using the expression (\ref{F21.9}) for the nucleus, we may write equation (%
\ref{F21.12}) as 
\begin{equation}
\label{F21.14} K_{p(q)}(q+Q,q)=e^{\frac{i}{\hbar}\lim_{N \rightarrow \infty}
\sum_{n=0}^{N}{p(q+n\delta q)\delta q} }. 
\end{equation}
The sum in the exponent above is clearly an integral taken along one period $%
Q$ of the system displacement on phase-space or, mathematically, 
\begin{equation}
\label{F21.15} K_{p(q)}(q+Q,q)=e^{\frac{i}{\hbar}\oint{p(q)dq} }. 
\end{equation}

Due to the periodicity condition (\ref{F21.10}) we shall have 
\begin{equation}
\label{F21.16} K_{p(q)}(q+Q,q)= \pm 1, 
\end{equation}
such that 
\begin{equation}
\label{F21.17} \oint{p(q)dq}=\left\{ 
\begin{array}{l}
2n\pi \hbar =nh 
\mbox{ if } K_{p(q)}(q+Q,q)=+1 \\ 2n\pi \hbar + \pi \hbar=(n+1/2)h 
\mbox{ if } K_{p(q)}(q+Q,q)=-1 
\end{array}
\right. , 
\end{equation}
implying that the choice between the quantum number $n$ or $n+1/2$ is
related with the transformation properties of the amplitude with respect to
a translation by its period on configuration space%
\footnote{
It is, indeed, related with the type of periodicity we are dealing with.
If the movement is a libration it will have turning points which will
introduce the extra phase $\pi /2$ responsible for the $1/2$ above. If this
movement is a rotation it will not have these turning points associated with
it and the $1/2$ factor will be absent. In fact, if the domain of the
coordinate is finet we expect the $1/2$ factor to appear while if the
domain of validity of $q$ is infinite, this factor will not be present.}.

This result shows that the appearance of half-integral quantum numbers is
expected even within the `old' quantum theory, something that could not be
predicted using the historical development of this theory \cite{Pauling}

Therefore, it explains why some systems will be described by half-integral
quantum numbers associated with them (as the harmonic oscillator) and others
will not (as the hydrogen atom).

The conditions (36) are precisely those of Bohr-Sommerfefld for the
stability of periodic systems, as we wished to derive.

In this sense, the present section reveals the intimate connection, for
periodic systems, between the `old' quantum theory and that one considered
`contemporary', represented by the Schr\"odinger equation. Such a
connections was indeed expected if one does not intend to consider the very
similar results obtained by both theories (e.g. for the energy spectrum of
the hydrogen atom) as being `accidental'.

The important difference between both approaches is that, while the one
based on the Schr\"odinger equation refers to {\it ensembles}, that one
based on the Bohr-Sommerfeld rules applies to each periodic system composing
this {\it ensemble} and is capable of explaining these systems stability, as
individual constituents of the {\it ensemble} considered.

\section{Path Integrals}

The approach of the last section may be slightly modified to allow us to
find the relation existing between the Schr\"odinger formal apparatus and
the one related with Feynman's path integrals \cite{Feynman1}.

To attain this goal, we shall use in (\ref{F21.6}) 
\begin{equation}
\label{F21.18} p\delta q=p\frac{\delta q}{\delta t} \delta t, 
\end{equation}
where we take formally 
\begin{equation}
\label{F21.19} \frac{\delta q}{\delta t}=\dot{q}, 
\end{equation}
meaning that $\delta q$ is being taken along a specific trajectory of the
system%
\footnote{
We may take the variation of a trajectory in a more general way
\cite{Goldstein3} using the expression
$$
\Delta q = \delta q+\dot{q}\Delta t,
$$
where, here, $\delta q$ means that we take this variation {\it between}
distinct trajectories. As we wish to vary the time along {\it one}
trajectory, we shall make $\delta q=0$ and use the expression
(\ref{F21.19})---with the necessary notational alterations. It is important
to stress here that the interpretation of this more general variation is
the one in which we are representing the {\it same} real trajectory on the
configuration space, varying only the velocity with which the point $q(t)$
moves along it. Using this variation, it is possible to show that
$$
\Delta \int_{t_1}^{t_2} {p\dot{q}dt}=\Delta \int_{q(t_1)}^{q(t_2)} {pdq}=0,
$$
and is such that if the system is periodic {\it in time} and we take
the integral above for only one period, we recover the results of the
previous section. The last expression is known as the Least Action Principle.}%
.

We may now use 
\begin{equation}
\label{F21.20} \dot{q}p=L(q,\dot{q};t)-E, 
\end{equation}
where $L$ is the classical lagrangean function and $E$ is the energy (here
supposed constant) of the system considered. In this case, expression (\ref
{F21.6}) becomes 
\begin{equation}
\label{F21.21} \psi(q(t+\delta t/2))=\int{\ e^{\frac{i}{2\hbar}\left[L(q,
\dot{q};t)-E\right] \delta t} \phi(q,\dot{q};t) J\left(\frac{p}{\dot{q}}%
\right) d(\dot{q}) }, 
\end{equation}
where 
\begin{equation}
\label{F21.22} J\left(\frac{p}{\dot{q}}\right)=\frac{dp}{d\dot{q}} 
\end{equation}
is the jacobian of the transformation $(q,p) \rightarrow (q,\dot{q})$.

The nucleus of the, infinitesimal in time, transformation (\ref{F21.21}) is
given by 
\begin{equation}
\label{F21.23} K_{\dot{q}(t)} (t+\delta t,t)=J\left(\frac{p(t)}{\dot{q}(t)}%
\right) e^{\frac{i}{\hbar}\left[L(q(t),\dot{q}(t);t)-E\right]\delta t}, 
\end{equation}
such that the transformation between two times $t_a=0$ and $t_b=t$ may be
written 
\begin{equation}
\label{F21.24} K_{\dot{q}(t)} (t_b,t_a)=lim_{N\rightarrow \infty}
\prod_{n=1}^{N} {\ K_{\dot{q}(t+(n-1)\delta t)} (t+n\delta t,t+(n-1)\delta
t) }, 
\end{equation}
where 
\begin{equation}
\label{F21.25} N\delta t=t_b-t_a, 
\end{equation}
making it necessary to take the limit $N\rightarrow \infty$, since $\delta t$
is infinitesimal.

Regrouping the terms in (\ref{F21.24}) we may find the expression 
\begin{equation}
\label{F21.26} K_{\dot{q}(t)} (t_b,t_a)= \left[ \lim_{N\rightarrow \infty}
\prod_{n=1}^{N}{J\left(\frac{p(t_n)}{\dot{q}(t_n)}\right)} \right] e^{\frac{i
}{\hbar}\lim_{N\rightarrow \infty}\sum[L(q(t_n),\dot{q}(t_n);t)-E] \delta
t}, 
\end{equation}
where we put 
\begin{equation}
\label{F21.26a} t_n=t+(n-1)\delta t, 
\end{equation}
and supposed that each limit exists. In the appropriate limit, we get 
\begin{equation}
\label{F21.27} K_{\dot{q}(t)} (t_b,t_a)=A e^{\frac{i}{\hbar}\int{L(q,\dot{q}%
;t)dt}} e^{-\frac{i}{\hbar}E(t_b-t_a)}, 
\end{equation}
if we put 
\begin{equation}
\label{F21.28} A=\lim_{N\rightarrow \infty} \prod_{n=1}^{N}{J\left(\frac{%
p(t_n)}{\dot{q}(t_n)}\right)} . 
\end{equation}

Now, writing 
\begin{equation}
\label{F21.29} S_{cl}[t_b,t_a]=\int_{t_a}^{t_b}{L(q,\dot{q};t)dt}, 
\end{equation}
we finally get the desired result 
\begin{equation}
\label{F21.30} K_{\dot{q}(t)} (t_b,t_a)=A e^{\frac{i}{\hbar}
S_{cl}[t_b,t_a]} e^{-\frac{i}{\hbar}E(t_b-t_a)}. 
\end{equation}

This last expression or its infinitesimal equivalent (\ref{F21.23}), is
exactly the one we obtain in the path integral approach of the quantum
formalism \cite{Feynman2}. The derivation method above has also the 
advantage of giving the mathematical expression of the constant $A$, as in 
(\ref{F21.28}).

This finally establishes the connection between the various methods of
quantization we have studied in this paper.

\section{Conclusions}

In this paper we have derived many important relations of the formal
apparatus of quantum mechanics. We will use this last section to make a
resum\'e of these relations, trying to fix the relevance of each one of them
within the theoretical approach proposed by ourselves since the first paper
of this series.

We may begin stressing the relevance of determining the physical
interpretation of the characteristic function (even for a specific example)
by fixing its relation with the probability density in configuration space.

Following the derived result, and making its generalization, we can consider
the quantum mechanical formalism as representing a statistical mechanics in
configuration space in which one studies the dynamic behavior of physical
systems which were infinitesimally dislocated from their mechanical
equilibrium situation. This establishes the relation between quantum
mechanics and historical equilibrium considerations, which are in the origin
of the formalism. Besides, it was possible to establish the connection
between the quantum formalism and thermodynamics, by means of the entropy
function.

Another important achievement was the derivation of the formal connection of
the quantization formalisms: one based on Schr\"odinger equation and
the other based on the quantization rules of Bohr-Sommerfeld; with the
important distinction made that the former refers to {\it ensembles} while
the latter to individual systems.

Finally, it was possible to link the Schr\"odinger formalism with the
Feynman path integral approach, fixing the interpretation of the
infinitesimal parameter once more.

\end{document}